\documentclass{PoS}
\usepackage{gensymb}
\usepackage{sidecap}
\usepackage{color}
\definecolor{refcolor}{rgb}{0.75,0.25,0.0}
\usepackage{cite}

\usepackage{lipsum}

\let\OLDthebibliography\thebibliography
\renewcommand\thebibliography[1]{
  \OLDthebibliography{#1}
  \setlength{\parskip}{0pt}
  \setlength{\itemsep}{0pt plus 0.3ex}
}

\title{The Extreme Universe Space Observatory on a Super-Pressure Balloon II Mission}

\ShortTitle{EUSO-SPB2}

\author{\speaker{L. Wiencke}$^a$ and A. Olinto$^b$
-- JEM-EUSO and POEMMA Collaborations \footnote{for collaboration list see PoS(ICRC2019)1177} \\ 
\llap{$^a$} Colorado School of Mines\\
\llap{$^b$} University of Chicago\\
E-mail: \email{lwiencke@mines.edu}}

\abstract{
The Extreme Universe Space Observatory on a Super Pressure Balloon II Mission (EUSO-SPB2) is a precursor for a next generation space observatory for multi-messenger astrophysics. The EUSO-SPB2 instrument will measure PeV and EeV-scale cosmic rays, optical backgrounds that could mimic tau neutrino interactions in the Earth's limb, and search for optical signatures consistent with the upward-going candidate events that the Antarctic Impulse Transient Antenna (ANITA) reported. The payload, now in the design and fabrication stage, features a pair of optical telescopes that have 1 meter diameter apertures. The Fluorescence Telescope will have sensitivity in the UV to target extensive air showers at EeV energies. The Cherenkov Telescope will have UV/VIS sensitivity to target Cherenkov emission from air showers at PeV energies. The planned launch date is 2022. We discuss the science and mission, and the current status.
}

\FullConference{36th International Cosmic Ray Conference -ICRC2019-\\
		July 24th - August 1st, 2019\\
		Madison, WI, U.S.A.}

\begin{document}
\section{Scientific Motivation for EUSO-SPB2}
Reaching $10^{20}$~eV (100~EeV), Ultra High Energy Cosmic rays (UHECRs) are the highest energy subatomic particles known to exist. The UHECR energy scale lies at least 5 orders of magnitude above that of the highest energy photons that have been measured from the cosmos. While the existence of UHECRs has been established for decades through measurements on the ground, their the sources and acceleration mechanisms remain unknown~\cite{Kotera:2011cp,Batista:2019,Anchordoqui:2018qom}. Measurements \cite{Auger:ExG} by the Pierre Auger Observatory indicate that the sources are not in our galaxy. 

Astrophysical PeV neutrinos, gamma rays, and UHECRs form a complementary trio of multi-messenger particles from some of the most extreme energetic environments of the universe. The recent intriguing multi-messenger observations reported by IceCube and collaborating gamma ray observatories of a flaring blazar\cite{IceCube:2018dnn} open an new window in neutrino astronomy with transients. 

\begin{figure} [h]
\centering
\includegraphics[width=4.0in,keepaspectratio]{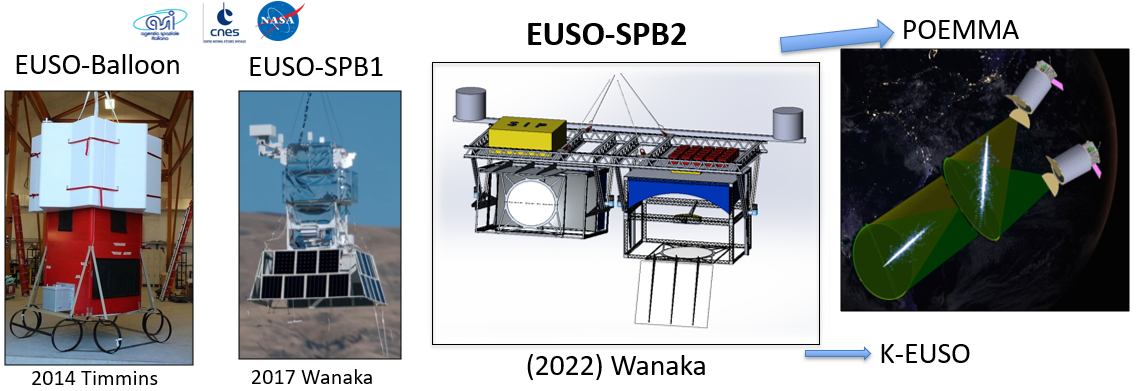} 
\caption{Evolution of the EUSO balloon payloads.}
\label{fig:EUSO-Balloons}
\end{figure}

EUSO-SPB2 is a scientific and technical sub-orbital altitude precursor for the Probe of Extreme Energy Multi-Messenger Astrophysics (POEMMA) \cite{POEMMA:2019}.  EUSO-SPB2 also provides R\&D toward the Klypve-Extreme Universe Space Observatory (K-EUSO) \cite{K-EUSO:2017} project. The evolution of EUSO balloon payloads is illustrated in Fig. \ref{fig:EUSO-Balloons}.

The main scientific goals of the POEMMA space mission are to discover the nature and origin of the highest energy particles in the universe\cite{P_UHECRS} and to discover neutrino emission above 20 PeV from extreme astrophysical transient sources\cite{P_ToO}. The list of possible target sources is extensive.

POEMMA recently completed a NASA-sponsored conceptual design. Briefly, POEMMA features two identical satellites that fly in formation to make coincident measurements from the dark side of the earth of light emission from Extensive Air Showers (EASs).  Each satellite is instrumented with Schmidt-based mirror optics with an entrance pupil diameter of 3.3~m. The focal surfaces are populated with Multianode Photomultiplier Tubes (MAPMTs) across most of their area and Silicone Photomultipliers (SiPMs) across region corresponding to the ``top'' part of the Field of View (FOV). For UHECRs, the MAPMTs on the two satellites measure the UV nitrogen fluorescence with 1 $\mu$s time bins in stereo constrain the geometric reconstruction of the EAS direction and location. The technique leads to an estimated energy resolution of about 20\% and depth of shower maximum resolution (Xmax) of 30 grams to enable measurements of UHECR energy, Xmax, and arrival direction. In turn, these parameters enable composition-enhanced anisotropy studies over the entire sky. POEMMA's sensitivity to astrophysical PeV neutrinos is maximized for earth-skimming $\nu_\tau$s from transient sources.  A PeV scale $\nu_\tau$ can interact in the earth's limb to produce a $\tau$ that exits the earth and decays in the atmosphere. The $\tau$ decay products initiate an EAS. The experimental signature in POEMMA is an upward-going flash of nearly parallel Cherenkov light from the EAS, observed simultaneously (after travel time correction), by both satellites in the SiPM parts of their focal surfaces. The SiPMs are digitized at 20~ns to increase the contrast of the fast Cherenkov flash above background. The satellites will be rotated temporarily from the nominal UHECR nadir pointing mode, to a new direction that centers the SiPM FOV on a ``Target of Opportunity'' transient source candidate that passes behind the earth's limb.

\section{EUSO-SPB2}
\begin{figure} [h]
\centering
\includegraphics[width=5.0in,keepaspectratio]{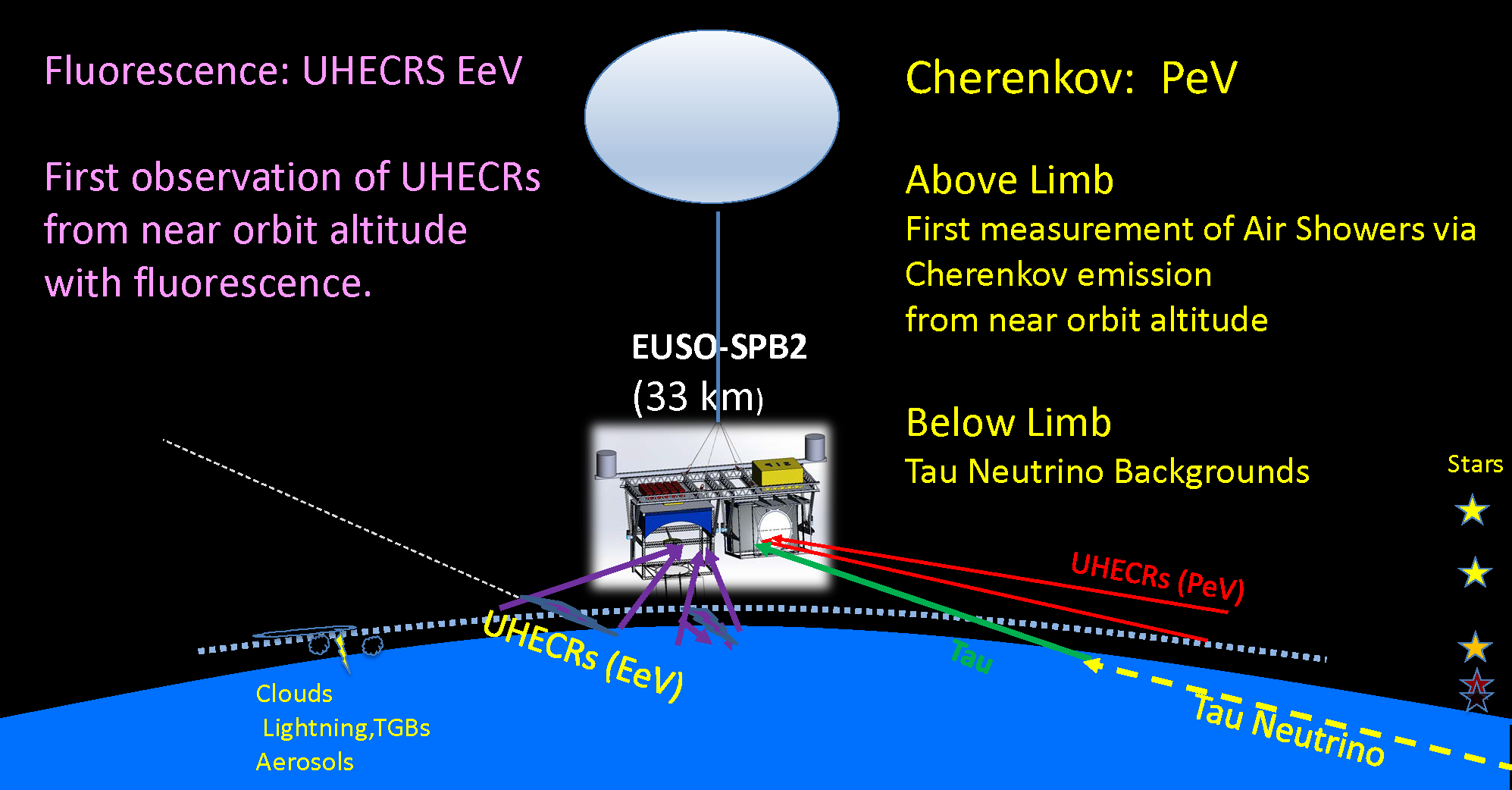}
\caption{A conceptual view of the EUSO-SPB2 configuration at float altitude. (not drawn to scale)}
\label{fig:Overview_Cartoon}
\end{figure}

The mission, methods and science goals of EUSO-SPB2 (Fig.\ref{fig:Overview_Cartoon}) are motivated by those of POEMMA. EUSO-SPB2 will fly two EAS telescopes at 33 km to make the first observations at near-orbit altitude of UHECRs with the UV fluorescence technique, PeV cosmic rays with the direct Cherenkov technique from near-orbit altitude, and measure $\nu_\tau$ optical background signatures by looking below the earth's limb. 

\begin{figure} [ht]
\centering
\includegraphics[width=5.8in,keepaspectratio]{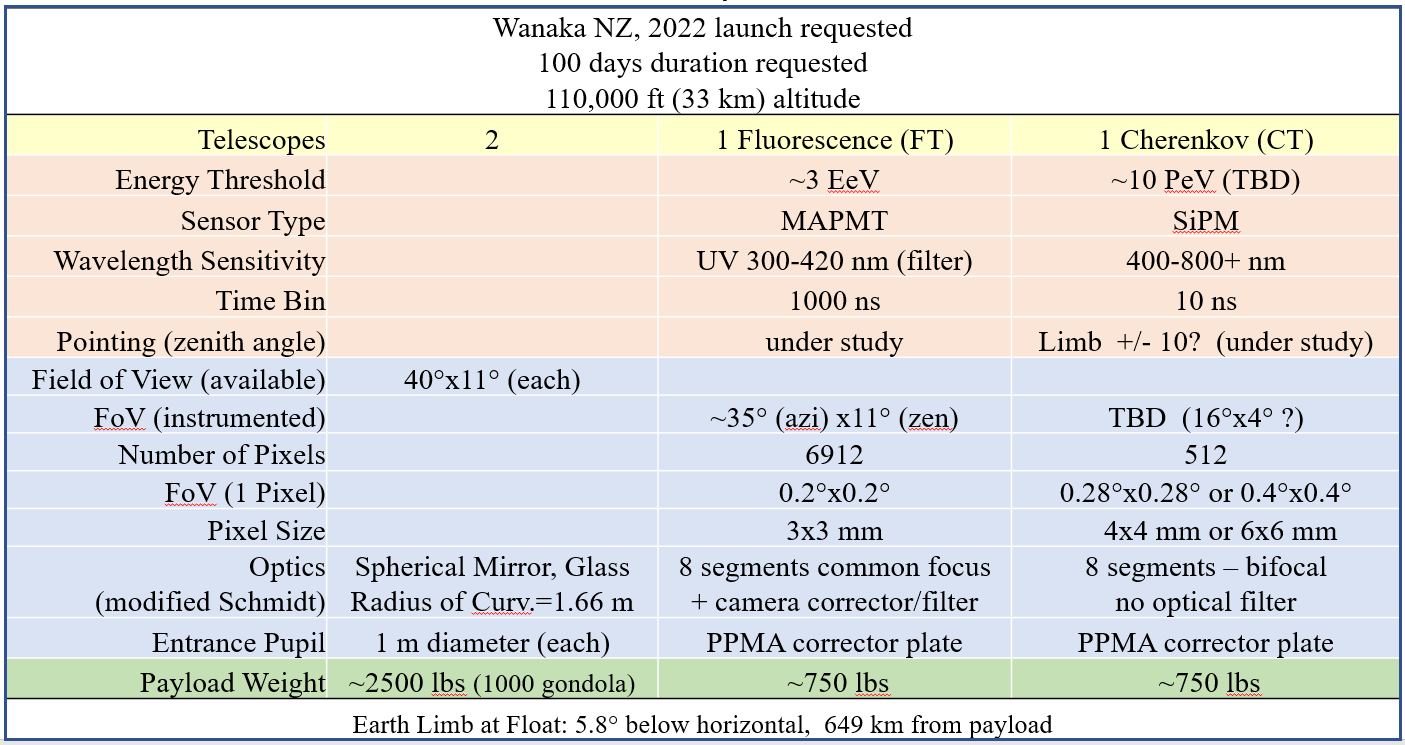} 
\caption{Selected parameters of the EUSO-SPB2 mission and instrument}
\label{fig:table_parameters}
\end{figure}

Key parameters of the EUSO-SPB2 payload, telescopes, and mission are listed in Fig.~\ref{fig:table_parameters}. Like POEMMA, EUSO-SPB2 will include downward viewing for UHECRs using MAPMTs and look near the earth limb with SiPMs. For logistical and technical simplicity, the two independent telescopes will fly on one payload with one telescope, denoted as the CT, configured for specifically for PeV Cherenkov measurements and the other, denoted as the FT, configured specifically for EeV UHECR fluorescence measurements. 

\begin{figure} [ht]
\centering
\includegraphics[width=4.5in,keepaspectratio]{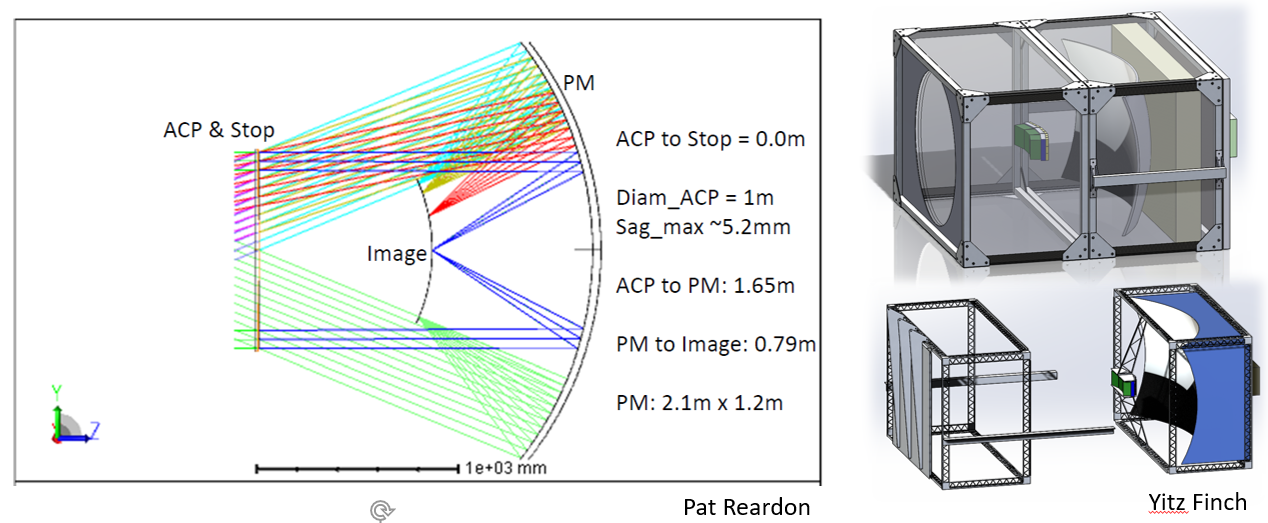} 
\caption{The optical design of EUSO-SPB2 (left panel) PM (Primary Mirror) APC (Aperture Corrector Plate) The telescope structure and camera location are shown in the right panel.}
\label{fig:Optics}
\end{figure}

The EUSO-SPB2 telescope optics (Fig. \ref{fig:Optics}) are a scaled down version of the POEMMA Schmidt-based optics with segmented spherical mirrors.
At 1~m diameter, the entrance pupil area of the EUSO-SPB2 telescopes is about 1/10 that of the POEMMA telescopes. However, EUSO-SPB2 will fly about 15 times closer to the ground.  Like POEMMA, the imaging requirements of  EUSO-SPB2 allow the optics to be designed as``light buckets'', ie far from diffraction-limited. The theoretical PSF (80\% of light spot) is about 1.5 mm diameter. A 3 mm pixel corresponds to a FOV of about 0.2\degree. The realized PSF for the 8 segment glass mirror of each telescope is expected to be 5 mm diameter after manufacturing and alignment precision and thermal effects. 

\begin{figure} [h]
\centering
\includegraphics[width=5.8in,keepaspectratio]{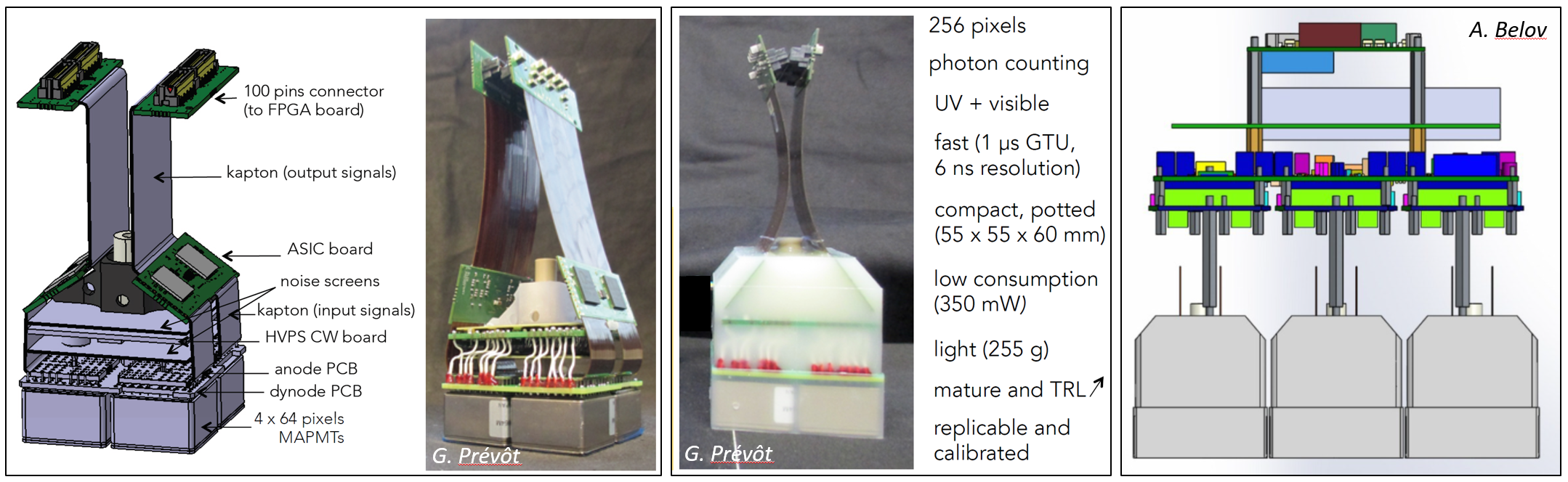} 
\caption{The 2x2 multi-anode photomultiplier assembly (256 channels) with integrated signal digization in photoelectron counts.  Left: Design and prototype.  Center: Prototype after potting.  Right: Side view of configuration of 2304 pixel array with trigger/event builder electronics and HV controls.}
\label{fig:EC}
\end{figure}

The Fluorescence telescope camera builds on the designs flown in the 2014 overnight EUSO-Balloon Mission that made the first observation of speed of light (laser) tracks by looking down\cite{Balloon-Laser}, the 2017 EUSO-SPB1 mission of opportunity \cite{SPB1-LW, SPB1-JE} that recorded optical backgrounds, instrument data and low energy cosmic rays  incident on the camera, but terminated prematurely in the Pacific Ocean, the EUSO-TA ground prototype \cite{EUSO-TA:2018ap} and the MINI-EUSO instrument \cite{MiniEUSO:2018asr} planned for launch to the ISS in 2019.  The EUSO-SPB2 FT camera features 256 pixel elementary cells (ECs) comprised of 4 MAPMTs (Fig.~\ref{fig:EC}). These ECs include signal integration to obtain photoelectron counts. The camera will include 27 ECs for a total of 6912 pixels.  One design challenge is accommodating the flat EC units to the curved image surface. A small corrector lens will be placed in front of each 3x3 EC array to preserve the PSF.  

\begin{figure} [h]
\centering
\includegraphics[width=4.5in,keepaspectratio]{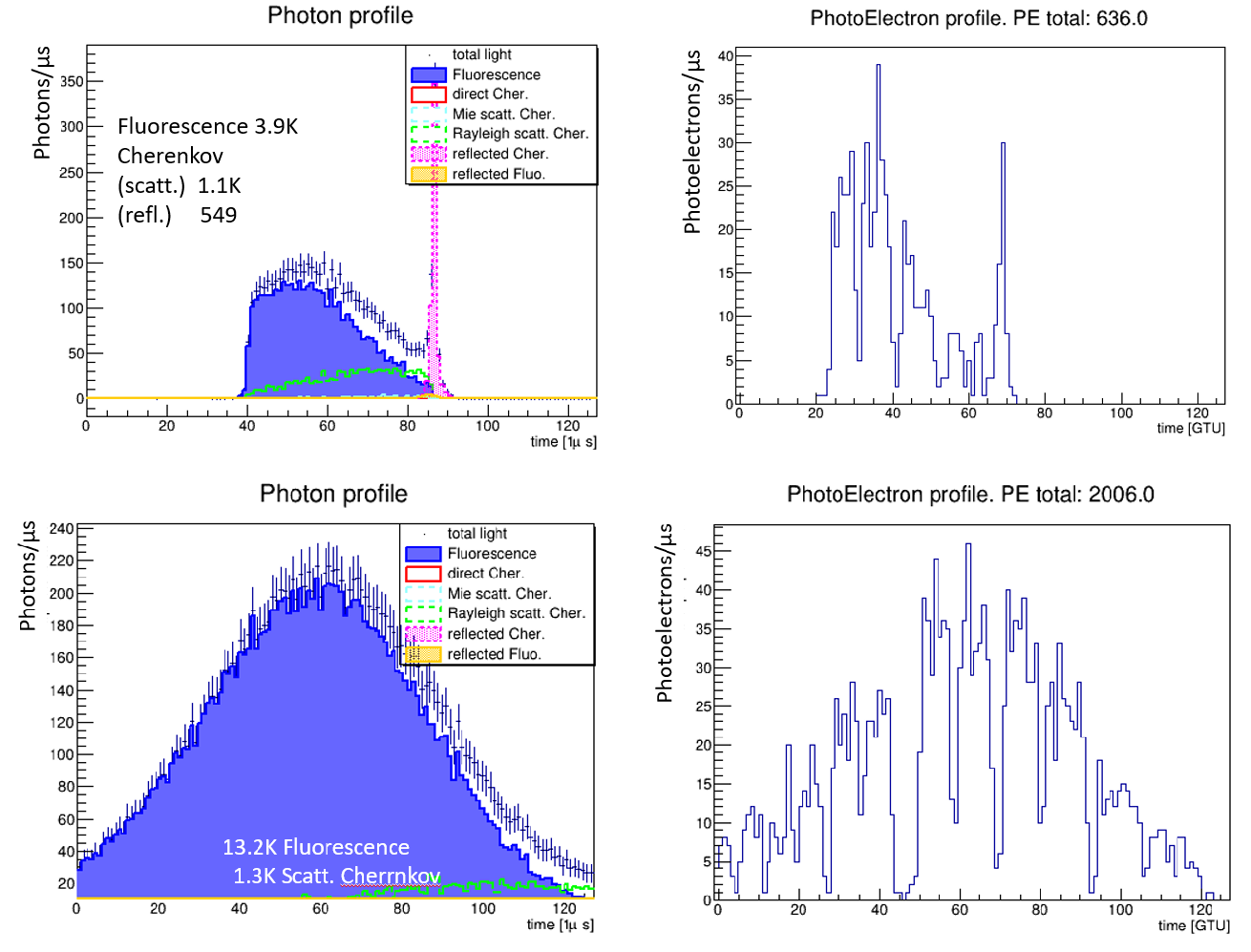} 
\caption{Simulated extensive air shower as recorded by EUSO-SPB2. Upper Left: a 3 EeV proton, 34\degree zenith angle. Light curve at telescope pupil with the FD tilted to nadir. Upper Right: Corresponding photoelectron counts, 1 $\mu$s time bin. Lower Left: a 10 EeV proton, 67\degree zenith angle. Light curve at telescope pupil with FD tilted 30\degree below horizontal.  Lower Right: Corresponding photoelectron counts, 1 $\mu$s time bin.}
\label{fig:FDevents}
\end{figure}

For the FT pointed in nadir mode, the EAS trigger rate currently estimated to be about 0.2/hr with a trigger threshold of about 1 EeV.  A 30\degree tilt angle below horizontal yields a 0.09/hr trigger rate with 10 EeV threshold. Examples of simulated EASs for which Xmax falls in the FT FOV are shown in Figs \ref{fig:FDevents}. 

The FT will also house a dual-band IR Camera camera system \cite{SPB2-IR:2019} designed to monitor the presence of clouds and measure cloud-top temperatures. 

\begin{figure} [h]
\centering
\includegraphics[width=5.8in,keepaspectratio]{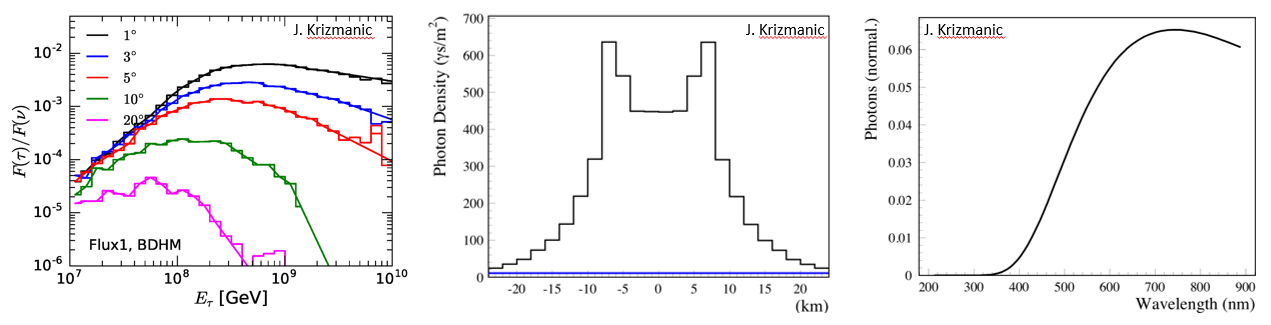} 
\caption{Preliminary simulations of Earth skimming $\nu_\tau$ signatures for the 33 km altitude of EUSO-SPB2. Left:  Ratio of tau emerging flux vs incident neutrino flux for different Earth emergence angles. Center: Cherenkov light distribution for a 100 PeV EAS 5 deg Earth emergence angle. The two peaks are separated by 3 deg at 33 km. Right: The corresponding Cherenkov light spectrum. }
\label{fig:tau_Cherenkov}
\end{figure}

A preliminary simulation of an expected Cherenkov signal \cite{nuSim:2019} signature of a Earth-skimming $\nu_\tau$ as seen from 33 km altitude is shown in Fig.~ \ref{fig:tau_Cherenkov}. Further simulations are in progress \cite{nuSim:2019}.

To provide a coincidence measurement of Cherenkov light spots from EASs, the CT mirror segments will be aligned in a bi-focal configuration. The bi-focal alignment is realized by rotating the upper 4 segments and the lower 4 segments relative to each other by a few PSF equivalents in azimuth. A parallel light pulse from outside the telescope will produce two lights spots on the focal surface and thus distinguish this trigger of interest from a single cluster of triggered pixels caused by a low energy cosmic primary striking the SiPM array.  Such triggers from direct cosmics were observed in the EUSO-SPB1 camera at float altitude.  The Cherenkov Telescope focal surface will be populated by 512 SiPMs. The final selection is being finalized, as is their arrangement on the image surface. Reference \cite{SPB2-CT}  details the camera development and status. 

Airglow MOMitor (AMON) and EUSO telescope darkness MONitor (EMON) detectors will be used as ancillary instruments.  AMON is a one pixel instrument \cite{Mackovjak:2019} that will measure the absolute intensities of the radiation from the limb and the atmosphere below. Two AMON instruments will operate in the same wavelength bands of the CT and FT to monitor airglow and night atmosphere radiation  backgrounds. The EMON instruments will be located inside the CT and FT telescopes to monitor darkness conditions during operation and testing.

Extensive laboratory and field testing is planned to characterize and calibrated the CT and FT prior to payload integration. Their PSFs and optical throughput will be measured for downward and horizontal orientations using a 1 m diameter optical test beam. The CT and FT will be transported to the Utah desert or a higher elevation location for testing with lasers and other light sources, building on the EUSO-SPB1 field campaign \cite{FC-Eser:2017}. This campaign will also include observation time to record EASs via Fluorescence and Cherenkov light.

\section{Mission}
A launch from NASA's Wanaka facility on a super pressure balloon is motivated by the opportunity for flight of up to 100 days. The 45\degree latitude, allows the FT and CT to collect data at night while the sky is dark. During the day solar panels recharge the batteries that power the instrument and other systems, including those for safe termination over land. NASA has been granted balloon over-flight and landing permissions routinely by the countries along this southern latitude. 

Wanaka lies under a stratospheric air circulation that forms each Fall and Spring. During the Southern Fall, this wide river of high thin air circles the Southern globe to the east at 50-150 km/h. A March/April launch window can insert the balloon into this circulation. The balloon and telescopes travel together in an apparent wind speed of essentially zero. The super pressure balloon is sealed. At float it expands to a maximum constrained volume of 18 million $ft^{3}$ with a slight over-pressure that remains positive at night to maintain the fixed volume. Thus the balloon floats at a nearly constant altitude. Absent technical issues, an SPB can remain aloft almost indefinitely.

\begin{figure} [h]
\centering
\includegraphics[width=4.7in,keepaspectratio]{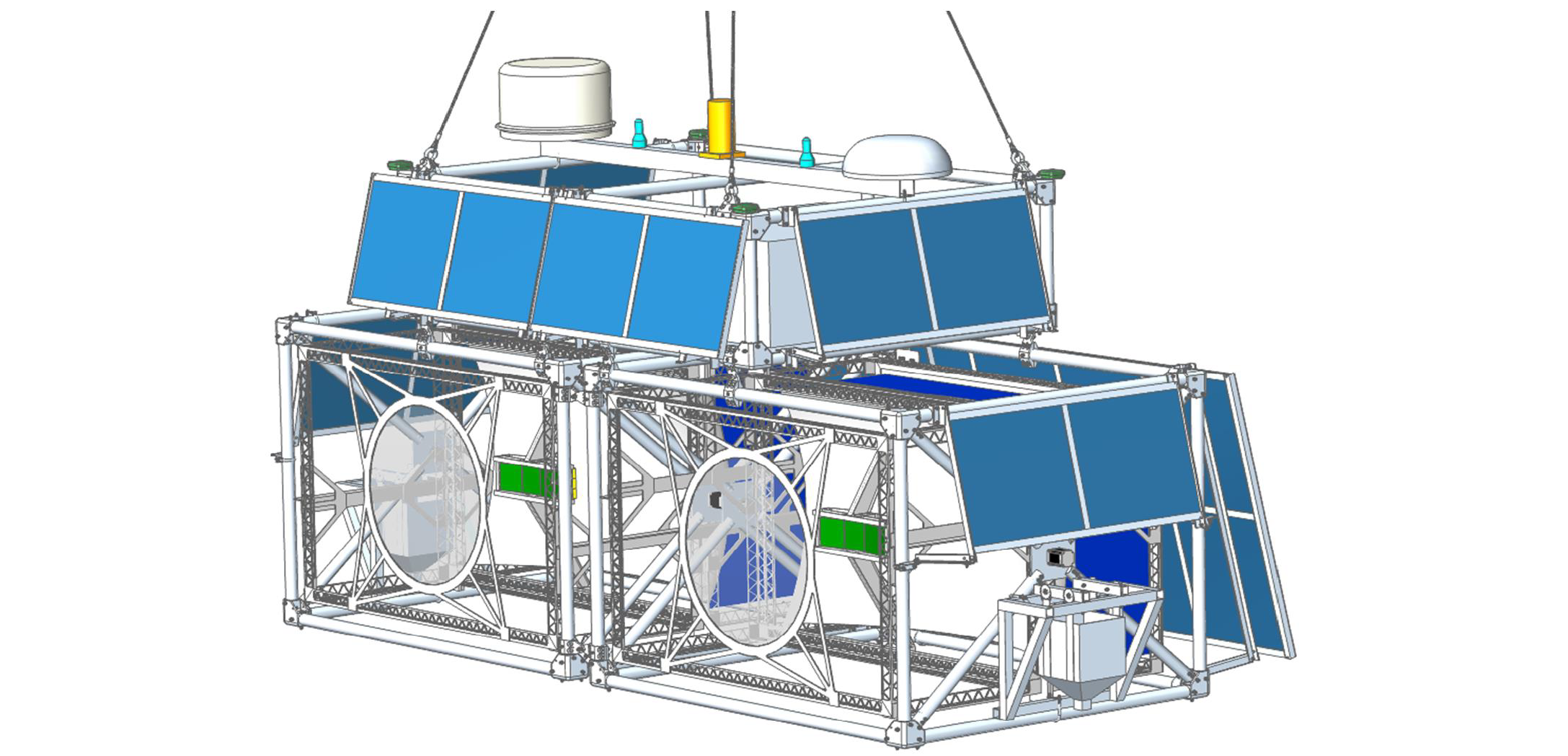} 
\caption{The EUSO-SPB2 payload including gondola structure shown in a premininary configuration, telescopes, solar panels, antennas, and ballast hoppers. }
\label{fig:gondola}
\end{figure}

To support the mission, the EUSO-SPB2 gondola (preliminary design: (Fig.~\ref{fig:gondola}) houses the CT, FT, the data and command telemetry systems, various antennas, a CPU, instrument monitoring systems, and two solar power systems. The gondola also houses systems for monitoring and operating the balloon and related systems, including ballast hoppers and flight termination control hardware, which are the responsibility of personnel from NASA's Columbia Scientific Balloon Facility.

The mission will focus on data collection during dark periods when the moon is down  for the best contrast for the optical signatures of scientific interest. The instrument solar panels will be mounted on the opposite side of the telescopes' apertures. A azimuth rotator with a sun tracker will turn the gondola so that this solar array tracks the sun. At night, the rotator will use feedback from a differential GPS system to control the azimuth direction. Nightly data taking will start with the telescopes opposite sunset in azimuth to catch the earliest dark backgrounds. Conversely, the nightly operation will terminate with the solar panels pointed toward the next sunrise. Instrument operations and monitoring will be shared between collaborators in the US, Japan and Europe to provide, during local daytime, control and monitoring of the telescopes 24/7. 

The CT will spend some observational time pointed above the -5.8\degree limb angle to record direct Cherenkov measurements from PeV cosmic rays. In addition to their first observation from near-orbit altitude, these events are expected to provide insitu estimates of the optics PSF and test the bifocal double-spot triggering algorithm. The bulk of the observing time will be devoted to measurements below the limb with the CT pointed in the directions from which upward tau-induced EASs are expected to originate. The zenith pointing direction(s) of the FT during the mission will be finalized once the simulations studies have been completed.  Searches for ANITA event-like candidates are under discussion.

This work was partially supported by 
NASA awards 11-APRA-0058, 80NSC18K0477,\\ 
80NSSC19K0627 in
the USA, Basic Science Interdisciplinary Research
Projects of RIKEN and JSPS KAKENHI Grant (JP17H02905, JP16H02426 and
JP16H16737), by the Italian Ministry of Foreign Affairs and International
Cooperation, by the Italian Space Agency through the ASI INFN agreement
n. 2017-8-H.0 and contract n. 2016-1-U.0, by  by the French space agency
CNES, by Slovak Academy of Sciences MVTS JEMEUSO and VEGA 
grant agency project
2/0132/17, by National Science Centre in Poland grant, 2017/27/B/ST9/02162)
Russia is supported by ROSCOSMOS and the Russian
Foundation for Basic Research Grant No 16-29-13065. Sweden is funded by the Olle Engkvist
Byggm\"astare Foundation.

\end{document}